# Chaotic behavior in Casimir oscillators: A case study for phase change materials

Fatemeh Tajik[1,2], Mehdi Sedighi[2], Mohammad Khorrami[1], Amir Ali Masoudi[1], George Palasantzas[2*]

[1]Department of Physics, Alzahra University, Tehran 1993891167, Iran

[2]Zernike Institute for Advanced Materials, University of Groningen, Nijenborgh 4, 9747 AG Groningen, The Netherlands

**Abstract**

Casimir forces between material surfaces at close proximity of less than 200 nm can lead to increased chaotic behavior of actuating devices depending on the strength of the Casimir interaction. We investigate these phenomena for phase change materials in torsional oscillators, where the amorphous to crystalline phase transitions lead to transitions between high and low Casimir force and torque states respectively, without material compositions. For a conservative system bifurcation curve and Poincare maps analysis show the absence of chaotic behavior but with the crystalline phase (high force/torque state) favoring more unstable behavior and stiction. However, for a non-conservative system chaotic behavior can occur introducing significant risk for stiction, which is again more pronounced for the crystalline phase. The latter illustrates the more general scenario that stronger Casimir forces and torques increase the possibility for chaotic behavior. The latter is making impossible to predict whether stiction or stable actuation will occur on a long term basis, and it is setting limitations in the design of micro/nano devices operating at short range nanoscale separations.

PACS numbers: 64.70.Nd, 85.85.+j, 12.20.Fv

___

*Corresponding author: g.palasantzas@rug.nl

# I. Introduction

Nowadays, advancements in fabrication techniques has led to scaling down of micromechanical systems into the submicron length scales, which open new areas of applications of the Casimir effect [1-7]. This is because micro/nano electromechanical systems (MEMS/NEMS) have surface areas large enough but gaps small enough for the Casimir force to play significant role. An example is a torsional actuator that is a kind of MEM with applications to torsional radio frequency (RF) switches, tunable torsional capacitors, torsional micro mirrors, and Casimir force measurements in the search of new forces beyond the standard model [1-4,8]. A simple torsional device (cantilever type) has two electrodes which one fixed and the other able to rotate around an axis [9]. The electrostatic and Casimir force can rotate the movable electrode towards the surface of the fixed electrode, and under certain conditions it can undergo jump-to-contact leading to permanent adhesion, a phenomenon known as stiction.

Although the Casimir force was predicted in 1948 [10], one must use the Lifshitz theory to compute the force between real dielectric materials [11]. This is accomplished by exploiting the fluctuation-dissipation theorem, which relates the dissipative properties of the plates (optical absorption by many microscopic dipoles) and the resulting electromagnetic field fluctuations that mediate the Casimir interaction between macroscopic bodies [11]. Since the optical properties of materials play crucial role on the Casimir force [12-14], it is anticipated to influence the actuation dynamics of MEMS. Indeed, it has been predicted that less conductive materials can enhance stable operation of MEMS in comparison to metal coated electrodes that yield higher Casimir forces [15]. In addition, there have been several investigations on Casimir torques [16-22] for possible applications on MEMS/NEMS. The genuine Casimir torque in periodic systems arise due to the broken rotational symmetry [16-18], while in optically anisotropic materials it originates from the misalignment between two optical axes [19-22]. Moreover, the actuation of MEMS can be influenced by mechanical Casimir torques originating from normal Casimir forces [23-27].

Furthermore, the magnitude of the Casimir force, and consequently the corresponding mechanical Casimir torque, can be modulated using, for example, the amorphous and crystalline phase transitions in phase change materials (PCMs) without composition changes [14]. Notably the similar possibilities were also explored using the metal-to-insulator phase transitions in hydrogen-switchable mirrors, and topological-insulator materials [28]. In any case the PCMs are renowned for their use in optical data storage (Blue-Rays, DVDs etc.) where they switch reversibly between the amorphous and crystalline phases [29]. Here we have chosen the AIST ($Ag_5In_5Sb_{60}Te_{30}$) PCM to perform our study, since we have measured the optical properties and the corresponding Casimir forces [14]. The amorphous phase of AIST is a semiconductor, while the crystalline phase shows closely metallic behavior [29], which is highly distinct from the amorphous state at low frequencies due to the high absorption of free carriers in the far-infrared (FAR-IR) spectrum [14]. Crystallization of the amorphous AIST has led up to ~25 % Casimir force contrast [14].

Therefore, PCMs offer a unique system to study how changes of the magnitude of the Casimir force and torque within the same system could affect the actuation dynamics of MEMS/NEMS. So far there is limited knowledge on how the Casimir forces/torques between actuating components at close proximity (typically less than 200 nm) can lead to chaotic behavior with changing strength of the force in relation also to the conduction properties of interacting materials. Surface roughness has been shown to strongly increase the Casimir force at separations less than 100 nm, and lead to chaotic behavior [30]. On the other hand, for flat surfaces, which are desirable in device application, this is also a possible scenario that has to be carefully investigated since Casimir forces are omnipresent. Hence, we will investigate here the occurrence of chaotic behavior in torsional oscillators when the amorphous to crystalline phase transitions lead to transitions between low and high Casimir force states respectively, though the conclusions have qualitatively general application for any material that is used in actuation of micro/nano devices.

## II. Theory of actuation system

The equation of motion for the torsional system (Fig. 1), where the fixed and rotatable plates are assumed to be coated with gold (Au) and AIST PCM respectively [14], is given by

$$I_0 \frac{d^2\theta}{dt^2} + \varepsilon\, I_0 \frac{\omega_0}{Q} \frac{d\theta}{dt} = \tau_{res} + \tau_{elec} + \tau_{Cas} + \varepsilon\, \tau_0 \cos(\omega t) \qquad (1)$$

where $I_0$ is the rotation inertia moment of the rotating plate. The conservative case corresponds to ε=0 and system quality factor Q=∞ (in practice $Q \geq 10^4$), while the non-conservative forced oscillation with dissipation to ε=1. The mechanical Casimir torque $\tau_{Cas}$ is given by [25]

$$\tau_{Cas} = \int_0^{L_x} r F_{Cas}(d') L_y\, dr \qquad (2)$$

where $F_{Cas}(d)$ is the Casimir force (See Supplemental Material [31] for Casimir force and dielectric function extrapolations in Figs.1s and 2s, as well as the dependency of the Casimir torque on the torsional angle for both PCM states in Fig. 3s), $L_x$ and $L_y$ are the length and width of each of the plates respectively (with $L_x = L_y = 10$µm), and $d' = d - r \sin\theta$ with d the distance for parallel plates. The torsional angle θ, which is considered positive as the plates move closer to each other, and its sign are also indicated in the inset of Fig. 1 that shows the actuating system. We assume also d=200 nm so that the maximum torsional angle $\theta_0$ to remain small ($\theta_0 = d/L_x = 0.02 \ll 1$) in order to ignore also any buckling of the moving beam (assuming typical operation at 300 K). Moreover, the electrostatic torque $\tau_{elec}$ due to an applied potential $V_a$ is given by [14, 25]

$$\tau_{elec} = \frac{1}{2}\varepsilon_0 L_y (V_a - V_c)^2 \frac{1}{\sin^2(\theta)} [\ln\left(\frac{d - L_x \sin(\theta)}{d}\right) + \frac{L_x \sin(\theta)}{d - L_x \sin(\theta)}] \quad (3)$$

with $\varepsilon_0$ the permittivity of vacuum, and $V_c$ is the contact potential difference between Au and AIST ($V_c \sim 0.4$ V for both phases of AIST) [14]. In the following we will consider only the potential difference $V = V_a - V_c$ for the Casimir torque calculations, and we will ignore small variations of $V_c$ between the amorphous and crystalline phases (~25 mV [14]). Both the Casimir and electrostatic torques are counterbalanced by the restoring torque $\tau_{res} = -k\theta$ with k the torsional spring constant around the support point of the beam [32]. Finally, the term $I_0(\omega/Q)(d\theta/dt)$ in Eq. (1) is due to the energy dissipation of the oscillating beam with Q the quality factor. The frequency ω is assumed to be typical like in AFM cantilevers and MEMS [1-4, 33].

In order to investigate the actuation dynamics by taking into account the effect of PCM phase transitions, we introduce the bifurcation parameter $\delta_{Cas} = \tau_{Cas}^m / k\theta_0$ that represents the ratio of the minimal Casimir torque $\tau_{Cas}^m = \tau_{Cas}(\theta = 0)$ for the amorphous phase of AIST, and the maximum restoring torque $k\theta_0$ [34]. Equation (1) can be rewritten in a normalized form in terms of $\delta_{Cas}$, $\varphi = \theta/\theta_0$, and the bifurcation parameter of the electrostatic force $\delta_V = (\varepsilon_0 V^2 L_y L_x^3)/(2kd^3)$ [21, 31],

$$\frac{d^2\varphi}{dT^2} + \varepsilon \frac{1}{Q}\frac{d\varphi}{dT} = -\varphi + \delta_v \frac{1}{\varphi^2}\left[\ln(1-\varphi) + \frac{\varphi}{1-\varphi}\right] + \delta_{Cas}\left[\frac{\tau_{cas}}{\tau_{Cas}^m}\right] + \varepsilon \frac{\tau_0}{\tau_{res}^{Max}}\cos(\frac{\omega}{\omega_0}T) \quad (4)$$

with $I = I_0/k$ and $T = \omega_0 t$.

### III. Conservative system ($\varepsilon=0$ and $Q=\infty$)

The equilibrium points for conservative motion are obtained by the condition $\tau_{total} = \tau_{res} + \tau_{elec} + \tau_{Cas} = 0$, which yields

$$-\varphi + \delta_v \frac{1}{\varphi^2}\left[\ln(1-\varphi) + \frac{\varphi}{1-\varphi}\right] + \delta_{Cas}\left[\frac{\tau_{cas}}{\tau_{Cas}^m}\right] = 0. \quad (5)$$

Figure 1 shows plots of $\delta_{Cas}$ vs. $\varphi$ for both the amorphous and crystalline phases for $\delta_v=0$ or equivalently V=0 (for $\delta_v > 0$ see Supplemental Material [31] the bifurcation curves in Figs.4s and 5s). Similarly to the Casimir bifurcation diagrams in Fig. 1, the bifurcation parameter $\delta_v$ also shows sensitive dependence on the amorphous to crystalline phase transition. In both cases the bifurcation curves of the amorphous and crystalline phases are distinct around the maximum, where one approaches critical unstable behavior. In Fig. 1 the solid lines show the stable regions where the restoring torque $\tau_{res}$ is strong enough to ensure stable periodic motion. The dash lines indicate unstable regions where the moving beam undergoes stiction. When $\delta_{Cas} < \delta_{Cas}^{MAX}$ two equilibrium points exist. The equilibrium point closer to $\varphi = 0$ (solid line) is a stable center point, and the other one closer to $\varphi = 1$ (dashed line) is the unstable saddle point. The latter obeys the additional condition $d\tau_{total}/d\varphi = 0$, which yields

$$-1 + \delta_v \left[\frac{2\varphi - 3}{\varphi^2(1-\varphi)^2} + \frac{2\ln(1-\varphi)}{\varphi^3}\right] + \delta_{Cas} \frac{1}{\tau_{Cas}^m}\left(\frac{d\tau_{Cas}}{d\varphi}\right) = 0. \tag{6}$$

By increasing $\delta_{Cas}$ or weakening the restoring torque ($\delta_{Cas} \sim 1/k$), the distance between the equilibrium points decreases until $\delta_{Cas}$ reaches the maximum saddle point $\delta_{Cas}^{MAX}$. In fact, when $\delta_{Cas} \sim \delta_{Cas,C}^{MAX}$ for the crystalline phase, it is still $\delta_{Cas} < \delta_{Cas,A}^{MAX}$ for the amorphous phase ensuring the presence of two equilibrium points and increased possibility for stable motion. The situation is qualitatively similar in presence of an electrostatic force (see Supplemental Material [31] the bifurcation curves in Fig.5s and Fig.6s). When the applied voltage increases $\delta_{Cas}^{MAX}$ decreases for both PCM phases. As a result, since the electrostatic force is attractive, the device would require higher restoring torque to ensure stable operation.

Besides the bifurcation diagrams, the sensitive dependence of the actuation dynamics on the PCM phase transition is reflected by the Poincare maps $d\varphi/dt$ vs. $\varphi$ in Fig. 2 [35]. The homoclinic orbit separates unstable motion (leading to stiction within one period, Fig. 3) from the periodic closed orbits around the stable center point. Since the distance between these two critical points is larger in the amorphous phase (see Supplemental Material [31] the phase portraits in Fig.7s), a torsional MEM can perform stable operation over a larger range of torsional angles. The orbit size in the crystalline phase is larger (Fig. 3a) because the moving plate approaches closer the fixed plate. With increasing $\delta_{Cas}$, the orbit breaks faster open for the crystalline phase leading to stiction (Fig. 3), while for the amorphous phase there is still periodic motion. Therefore, the amorphous phase can ensure better device stability without any significant differences in electrostatic contributions (due to some difference in $V_c$ [11]) from the crystalline phase.

Moreover, if one introduces some dissipation into the autonomous oscillating system via a finite quality factor Q, then dissipative motion can prevent stiction also for the crystalline phase despite the stronger Casimir torque (Fig. 3b; see Supplemental Material [31] the Phase portraits in Fig.8s). In any case, because the homoclinic orbit separates qualitatively different (stable/unstable) solutions, as the Poincare maps show in Fig.2, it precludes the possibility of chaotic motion or equivalently sensitive dependence on the initial conditions [30, 35]. A chaotic oscillator can have qualitatively different solutions for an arbitrarily small difference in the initial conditions. As a result the conservative oscillating system provides an essential reference for the study of forced oscillations induced by an external applied forces and torque treated as a perturbative correction on the conservative system.

## IV. Non-conservative system ($\varepsilon=1$ and $Q<\infty$)

Here we performed calculations to investigate the existence of chaotic behavior of the torsional system undergoing forced oscillation via an applied external torque $\tau_o \cos(\omega t)$ [30]. Chaotic behavior occurs if the separatrix (homoclinic orbit) of the conservative system splits, which it can be answered by the so-called Melnikov function and Poincare map analysis [30, 35]. If we define the homoclinic solution of the conservative system as $\varphi_{hom}^C(T)$, then the Melnikov function for the torsional system ($\varepsilon = 1$) is given by [30, 35]

$$M(T_0) = \frac{1}{Q}\int_{-\infty}^{+\infty}\left(\frac{d\varphi_{hom}^C(T)}{dT}\right)^2 dT + \frac{\tau_0}{\tau_{res}^{MAX}} \int_{-\infty}^{+\infty} \frac{d\varphi_{hom}^C(T)}{dT} \cos\left[\frac{\omega}{\omega_0}(T + T_0)\right] dT \qquad (7)$$

The separatrix splits if the Melnikov function has simple zeros so that $M(T_0) = 0$ and $M'(T_0) \neq 0$. If $M(T_0)$ has no zeros, the motion will not be chaotic. The conditions of nonsimple zeros,

$M(T_0) = 0$ and $M'(T_0) = 0$ gives the threshold condition for chaotic motion [30, 35]. If we define

$$\mu_{hom}^C = \int_{-\infty}^{+\infty} \left(\frac{d\varphi_{hom}^C(T)}{dT}\right)^2 dT \text{ and } \beta(\omega) = \left|H\left[\text{Re}\left(F\left\{\frac{d\varphi_{hom}^C(T)}{dT}\right\}\right)\right]\right|, \quad (8)$$

then the threshold condition for chaotic motion $\alpha = \beta(\omega)/\mu_{hom}^C$ with $\alpha = (1/Q)(\tau_0/\tau_{res}^{MAX})^{-1} = \gamma\omega_0 \theta_0/\tau_0$ obtains the form

$$\alpha = \frac{\gamma\omega_0 \theta_0}{\tau_0} = \left|H\left[\text{Re}\left(F\left\{\frac{d\varphi_{hom}^C(T)}{dT}\right\}\right)\right]\right| / \int_{-\infty}^{+\infty} \left(\frac{d\varphi_{hom}^C(T)}{dT}\right)^2 dT \quad (9)$$

with $\gamma = I\omega_o/Q$, and $H[...]$ denoting the Hilbert transform [30, 35]. Figure 4 shows the threshold curves $\alpha = \gamma\omega_0 \theta_0/\tau_0$ vs. driving frequency ratio $\omega/\omega_o$. For large values of $\alpha$ (above the curve) the dissipation dominates the driving torque ($\alpha \sim \gamma/\tau_0$) leading to regular motion, which asymptotically approaches the stable periodic orbit of the conservative system. However, for parameter values below the curve, the splitting of the separatrix leads to chaotic motion. Clearly for the crystalline state, which gives to stronger Casimir torques, chaotic motion is more likely to occur.

Since we study the occurrence of chaotic motion in terms of the sensitive dependence of the motion on its initial conditions, we present in Fig. 5 Poincare maps for different values of the threshold parameter $\alpha$. When chaotic motion occurs (with decreasing value of $\alpha$) there is a region of initial conditions where the distinction between qualitatively different solutions is unclear. If we compare with Fig. 2, where chaotic motion does not occur, the latter implies that for chaotic motion there is no a simple smooth boundary between the *red* and the *dark-blue* regions. As a result, if the motion is chaotic then stiction can take place after several periods affecting the long-term stability of the device. Therefore, chaotic behavior introduces significant risk for stiction and this more prominent to occur for the more conductive crystalline PCM. In more general, as the Casimir force/torque increases the possibility for chaotic behavior increases and practically it could be impossible to predict whether stiction or stable actuation will occur on a long term basis.

## V. Conclusions

In conclusion, Casimir forces and torques between actuating components at close proximity, typically less than 200 nm, can lead to increased chaotic behavior with increasing strength of the non-linear in nature Casimir interaction. We have illustrated these phenomena in torsional oscillators undergoing both conservative and non-conservative motion, where the amorphous to crystalline phase transitions in phase change materials lead to transitions between high and low Casimir force/torque states respectively. The occurrence of chaotic behavior introduces significant risk for stiction, and this more prominent for the more conductive crystalline phase that generates stronger Casimir forces and torques. In addition, this is also the case for conservative motion, where chaotic behavior is absent, that the crystalline phase is again more luckily to lead to stiction.

For the particular case of PCMs, our study shows that these materials can offer a versatile way to control motion by using both phases of the PCMs and controlled energy dissipation during device actuation. Furthermore, our analysis has general character in the sense that as the Casimir force/torque increases the possibility for chaotic behavior increases, and practically it could impossible to predict whether stiction or stable actuation will occur on a long term basis. The latter has serious implications because Casimir forces are omnipresent, and one must be very careful in choosing the proper conductivity materials in the design of micro/nano devices actuating at nanoscale separations.

## Acknowledgements

GP acknowledges support from the Zernike Institute of Advanced Materials, University of Groningen. FT, MK, and AAM acknowledge support from the Department of Physics at Alzahra University.

**Figure 1** Bifurcation diagrams $\delta_{Cas}$ vs. $\varphi$ for $\delta_v = 0$. The solid and dashed lines represent the stable and unstable points respectively. The inset shows the schematic of the torsional system.

**Figure 2** Poincare maps $d\varphi/dt$ vs. $\varphi$ ($\delta_{Cas} = 0.1$, $\delta_v = 0$) of the conservative system ($\varepsilon=0$) for amorphous and crystalline PCM phases. For the calculations we used 150×150 initial conditions ($\varphi$, $d\varphi/dt$). The red region shows that initial condition for which the torsional device shows stable motion after 100 oscillations with the natural frequency $\omega_0$. The homoclinic orbit separates sharply stable and unstable solutions prohibiting chaotic behavior.

**Figure 3** Phase portraits $d\varphi/dt$ vs. $\varphi$ for $\delta_{Cas} = 0.1$, $\delta_v = 0$ and Q=∞. (a) Similar plot for smaller $\delta_{Cas} = 0.09$ where only stable motion takes place for both PCM phases. (b) Phase portraits for $\delta_{Cas} = 0.1$, $\delta_v = 0$, and finite damping contributing with Q=500.

**Figure 4** Threshold curve $\alpha$ ($= \gamma\omega_0 \theta_0/\tau_0$) vs. driving frequency $\omega/\omega_o$ (with $\omega_o$ the natural frequency of the system) for the amorphous and crystalline states. The area bellow the curve corresponds to chaotic motion.

**Figure 5** Poincare maps $d\varphi/dt$ vs. $\varphi$ ($\delta_{Cas} = 0.1$, $\delta_v = 0$) of the non-conservative system ($\varepsilon=1$) for amorphous (left column) and crystalline (right column) PCM phases. For the calculations we used 150×150 initial conditions ($\varphi$, $d\varphi/dt$). The red region shows that initial condition for which the torsional device shows stable motion after 100 oscillations with oscillating frequency

$\omega/\omega_0 = 0.8$. With decreasing $\alpha$ the chaotic behavior increases, and the area of stable motion shrinks more for the crystalline (high force/torque) phase.

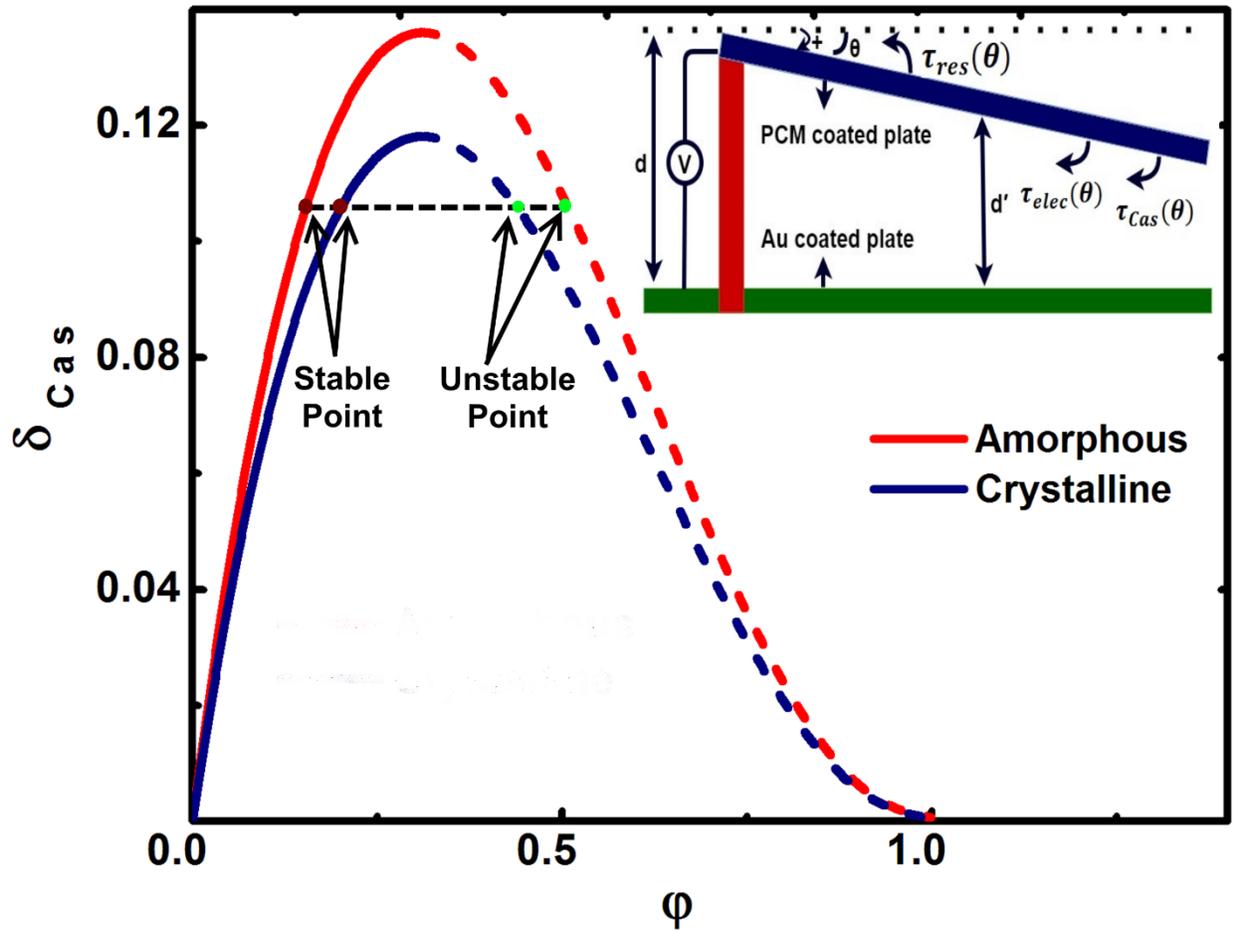

**Figure 1**

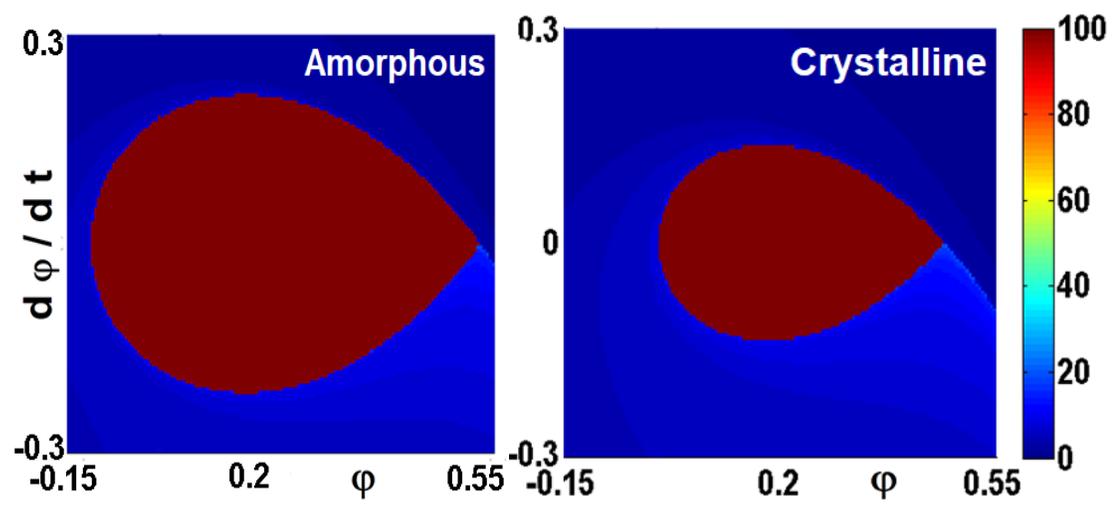

**Figure 2**

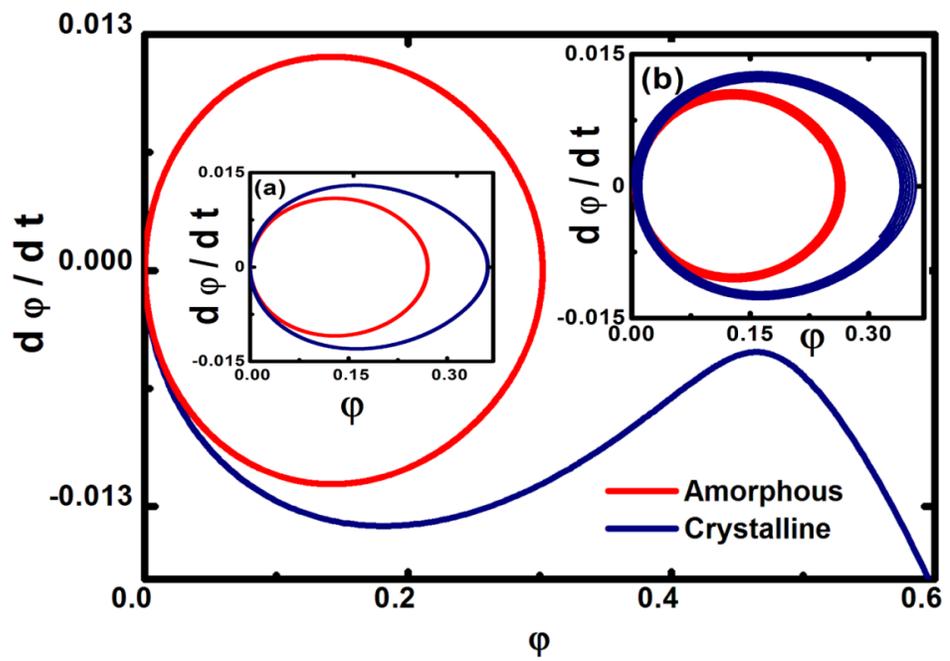

**Figure 3**

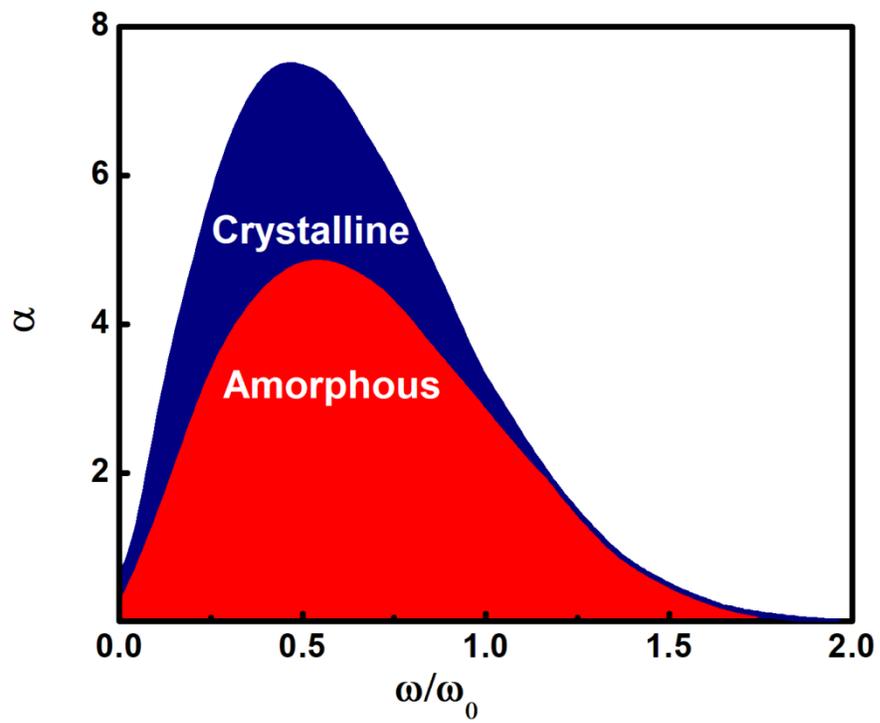

**Figure 4**

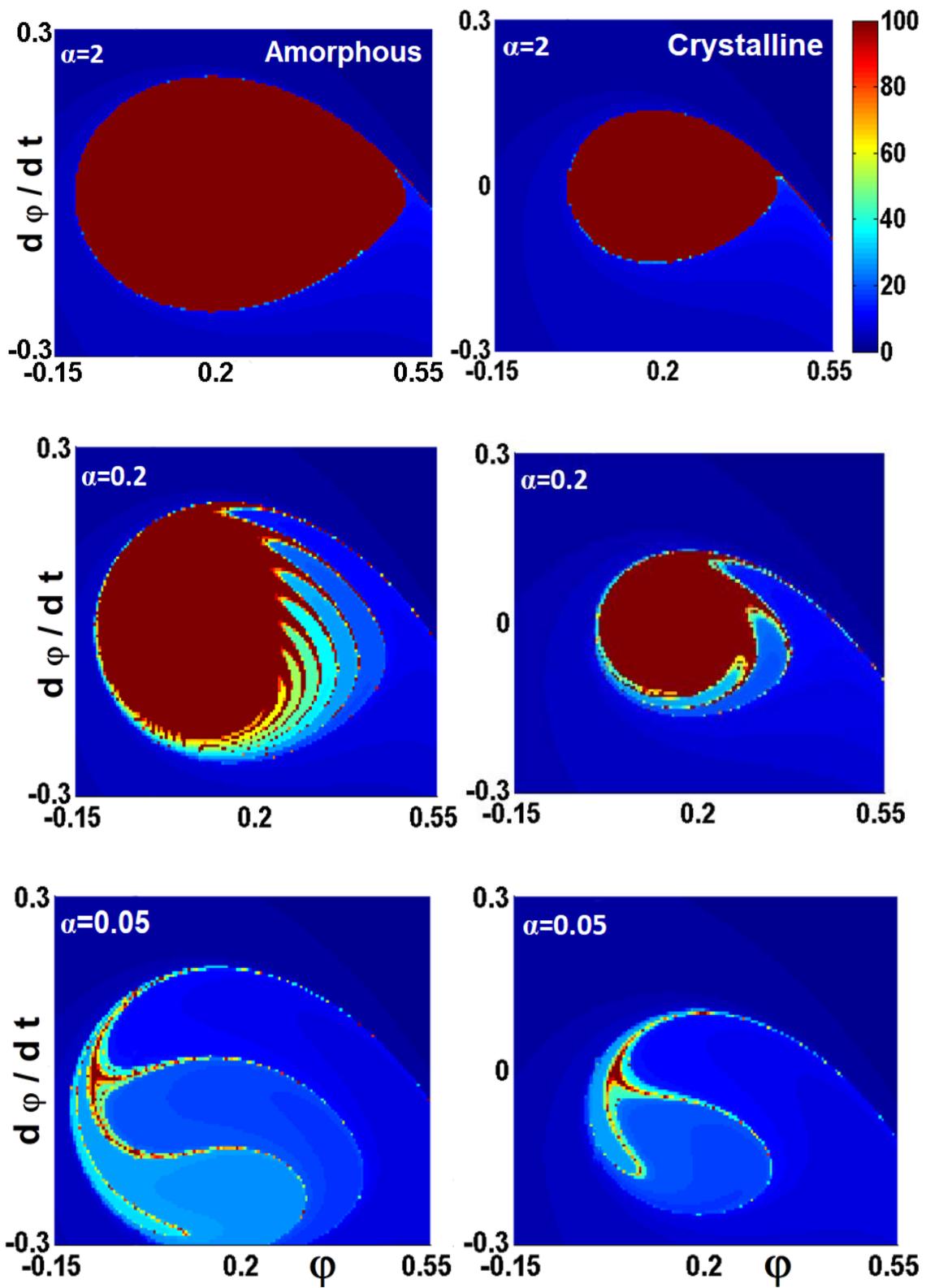

Figure 5

**Supplemental material**

**Chaotic behavior in Casimir oscillators: A case study for phase change materials**

Fatemeh Tajik[1,2], Mehdi Sedighi[2], Mohammad Khorrami[1], Amir Ali Masoudi[1], George Palasantzas[2]

[1]Department of Physics, Alzahra University, Tehran 1993891167, Iran
[2]Zernike Institute for Advanced Materials, University of Groningen, Nijenborgh 4, 9747 AG Groningen, The Netherlands

## 1. Casimir force and Dielectric function of PCM materials with extrapolations

The Casimir force $F_{Cas}(d)$ in Eq.(2) is given by (s1)

$$F_{Cas}(d) = \frac{k_B T}{\pi} \sum_{l=0}^{\prime} \sum_{\nu=TE,TM} \int_0^\infty dk_\perp \, k_\perp \, k_0 \frac{r_\nu^{(1)} r_\nu^{(2)} \exp(-2k_0 d)}{1 - r_\nu^{(1)} r_\nu^{(2)} \exp(-2k_0 d)}. \tag{S1}$$

The prime in first summation indicates that the term corresponding to $l = 0$ should be multiplied with a factor $1/2$. The Fresnel reflection coefficients are given by $r_{TE}^{(i)} = (k_0 - k_i)/(k_0 + k_i)$ and $r_{TM}^{(i)} = (\varepsilon_i k_0 - \varepsilon_0 k_i)/(\varepsilon_i k_0 + \varepsilon_0 k_i)$ for the transvers electric and magnetic field polarizations respectively. $k_i(i = 0,1,2) = \sqrt{\varepsilon_i(i\xi_l) + k_\perp^2}$ represents the out-off plane wave vector in the gap between the plates ($k_0$), and in each of the interacting plates ($k_{i=(1,2)}$), as well as $k_\perp$ is the in-plane wave vector.

Furthermore, $\varepsilon(i\xi)$ is the dielectric function evaluated at imaginary frequencies, which is the necessary input for calculating the Casimir force between real materials using Lifshitz theory. The latter is given by [s1-s4]

$$\varepsilon(i\xi) = 1 + \frac{2}{\pi} \int_0^\infty \frac{\omega \, \varepsilon''(\omega)}{\omega^2 + \xi^2} d\omega. \tag{S2}$$

For the calculation of the integral in Eq. (S1) one needs the measured data for the imaginary part $\varepsilon''(\omega)$ (Fig. 1s) of the frequency dependent dielectric function $\varepsilon(\omega)$. The AIST PCM was optically characterized by ellipsometry over a wide range of frequencies at J. A.Woollam Co.: VUV-VASE (0.5–9.34 eV) and IR-VASE (0.03–0.5 eV)) [s3, s4].

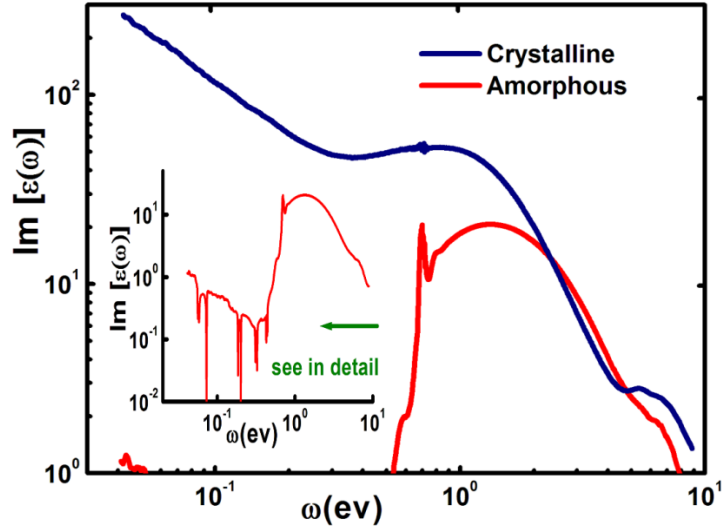

**Figure 1s** Imaginary part $\varepsilon''(\omega)$ of the frequency-dependent dielectric function for both phases of AIST [s3, s4].

The experimental data for the imaginary part of dielectric function cover only a limit range of frequencies $\omega_1 (= 0.03 \text{ ev}) < \omega < \omega_2 (= 8.9 \text{ ev})$. Therefore, for the low optical frequencies ($\omega < \omega_1$) we extrapolated using the Drude model for the crystalline phase [s3, s4]

$$\varepsilon''_L(\omega) = \frac{\omega_p^2 \, \omega_\tau}{\omega \, (\omega^2 + \omega_\tau^2)}, \tag{S3}$$

where $\omega_p$ is the Plasma frequency, and $\omega_\tau$ is the relaxation frequency. For amorphous phase there is no any significant IR absorption and contribution to the Casimir force. For the crystalline phase the free carriers have small mean free paths (below ~3 nm) implying a significant value for $\omega_\tau$ [3]. Therefore, the extrapolation via the Drude model in Eq.(S3), since $\omega \ll \omega_\tau$, obtains the form

$$\varepsilon''_L(\omega) = \frac{\omega_p^2}{\omega \omega_\tau}, \tag{S4}$$

where one can determine from the optical data directly the ratio $\omega_p^2/\omega_\tau$ [s4]. Furthermore, for the high optical frequencies ($\omega > \omega_2$) we extrapolated using for both PCM phases [s5]

$$\varepsilon''_H(\omega) = \frac{A}{\omega^3}. \tag{S5}$$

Finally, using Eqs. (S2)-(S5) $\varepsilon(i\xi)$ is given for both phases by

$$\varepsilon(i\xi)_A = 1 + \frac{2}{\pi} \int_{\omega_1}^{\omega_2} \frac{\omega \, \varepsilon''_{\exp}(\omega)}{\omega^2 + \xi^2} \, d\omega + \Delta_H \varepsilon(i\xi) \tag{S6}$$

$$\varepsilon(i\xi)_C = 1 + \frac{2}{\pi} \int_{\omega_1}^{\omega_2} \frac{\omega \, \varepsilon''_{\exp}(\omega)}{\omega^2 + \xi^2} \, d\omega + \Delta_L \varepsilon(i\xi) + \Delta_H \varepsilon(i\xi) \tag{S7}$$

Equations (S6) and (S7) are the dielectric functions for the amorphous and crystalline phases respectively with

$$\Delta_L \varepsilon(i\xi) = \frac{2}{\pi}\int_0^{\omega_1} \frac{\omega\, \varepsilon''_L(\omega)}{\omega^2 + \xi^2}\, d\omega, \text{ and } \Delta_H \varepsilon(i\xi) = \frac{2}{\pi}\int_{\omega_2}^{\infty} \frac{\omega\, \varepsilon''_H(\omega)}{\omega^2 + \xi^2}\, d\omega. \tag{S8}$$

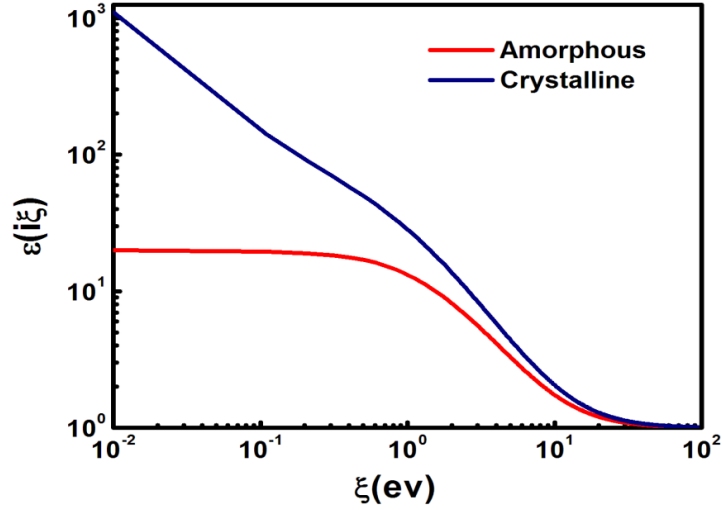

**Figure 2s** Dielectric functions at imaginary frequencies $\varepsilon(i\xi)$ for the AIST PCM.

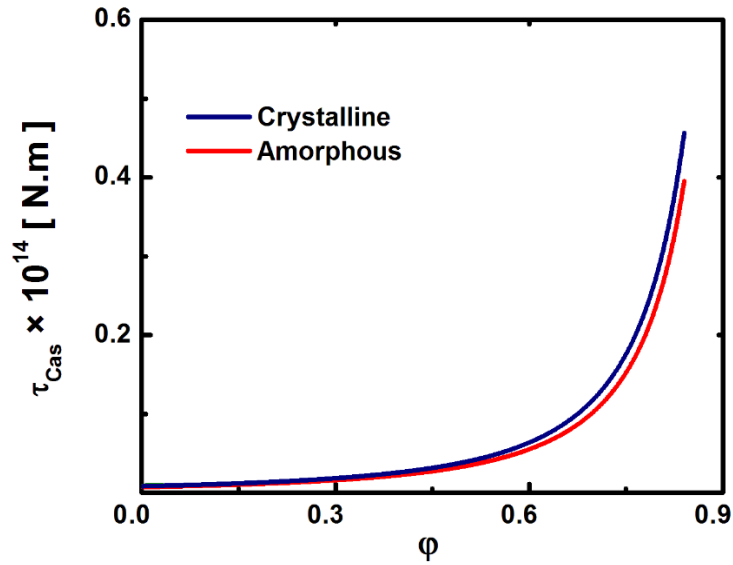

**Figure 3s** Casimir torques calculated for Au-PCM materials using as input the optical data from Fig. 2s.

## 2. Bifurcation curves in presence of applied voltage

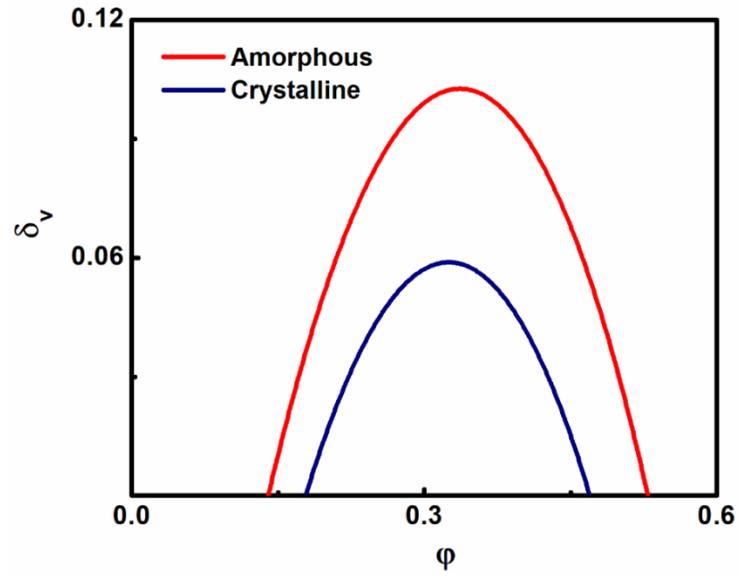

**Figure 4s** Bifurcation diagrams $\delta_v$ vs. $\varphi$ for $\delta_{Cas} = 0.1$.

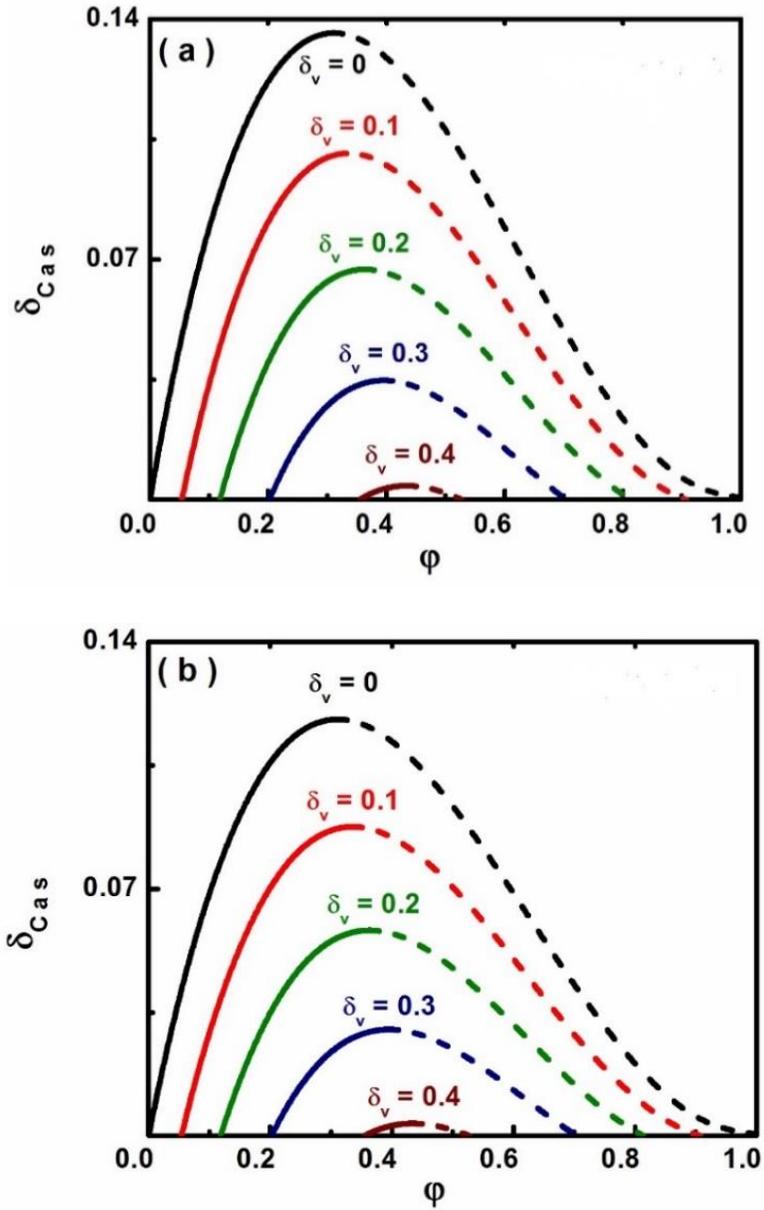

**Figure 5s** Bifurcation diagrams $\delta_{Cas}$ vs. $\varphi$ for different $\delta_v$. All points of the solid and dashed lines represent the stable and unstable points respectively in (a) amorphous and (b) crystalline phase. $\delta_{Cas}^{MAX}$ decreases in magnitude if one compares the amorphous and crystalline phases.

According to the diagram of the bifurcation parameter $\delta_v$, the maximum $\delta_v^{MAX}$ decreases similar to $\delta_{Cas}^{MAX}$. The range of bifurcation parameters to produce periodic motion ($0 < \delta_{Cas} < \delta_{Cas}^{MAX}$ and $\delta_v \geq 0$) is decreased during the amorphous to crystalline phase transition. Note that for $\delta_{Cas} > \delta_{Cas}^{MAX}$ there is no stability in the torsional device even in the absence of electrostatic torques ($\delta_v = 0$).

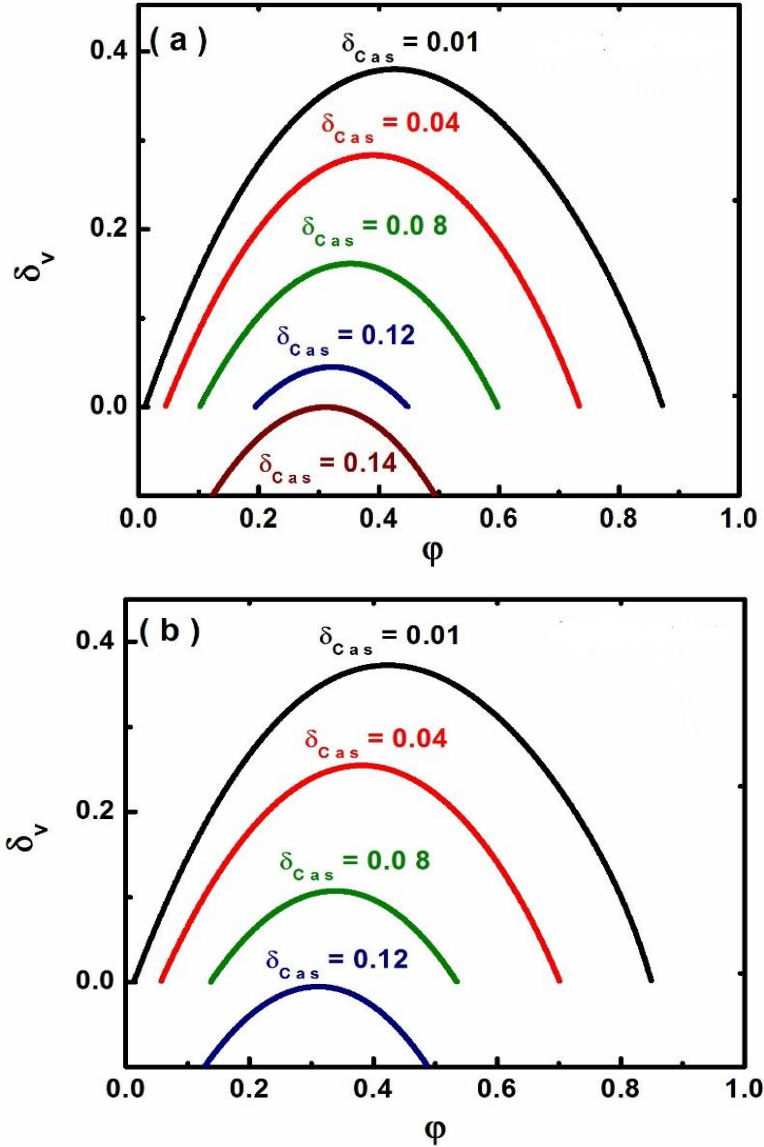

**Figure 6s** Variation of $\delta_v$ for different values of $\delta_{Cas}$ in (a) amorphous, and (b) crystalline phases. It can be clearly seen that for $\delta_{Cas} \leq 0.12$ we have $\delta_v \geq 0$. The latter means that for $\delta_{Cas} > 0.12$ there is no stability even without any voltage. For the amorphous phase the value of the critical $\delta_{Cas}$ is larger and a weaker restoring torque can lead to stable actuation.

## 3. Comparison of phase portrait between two phases

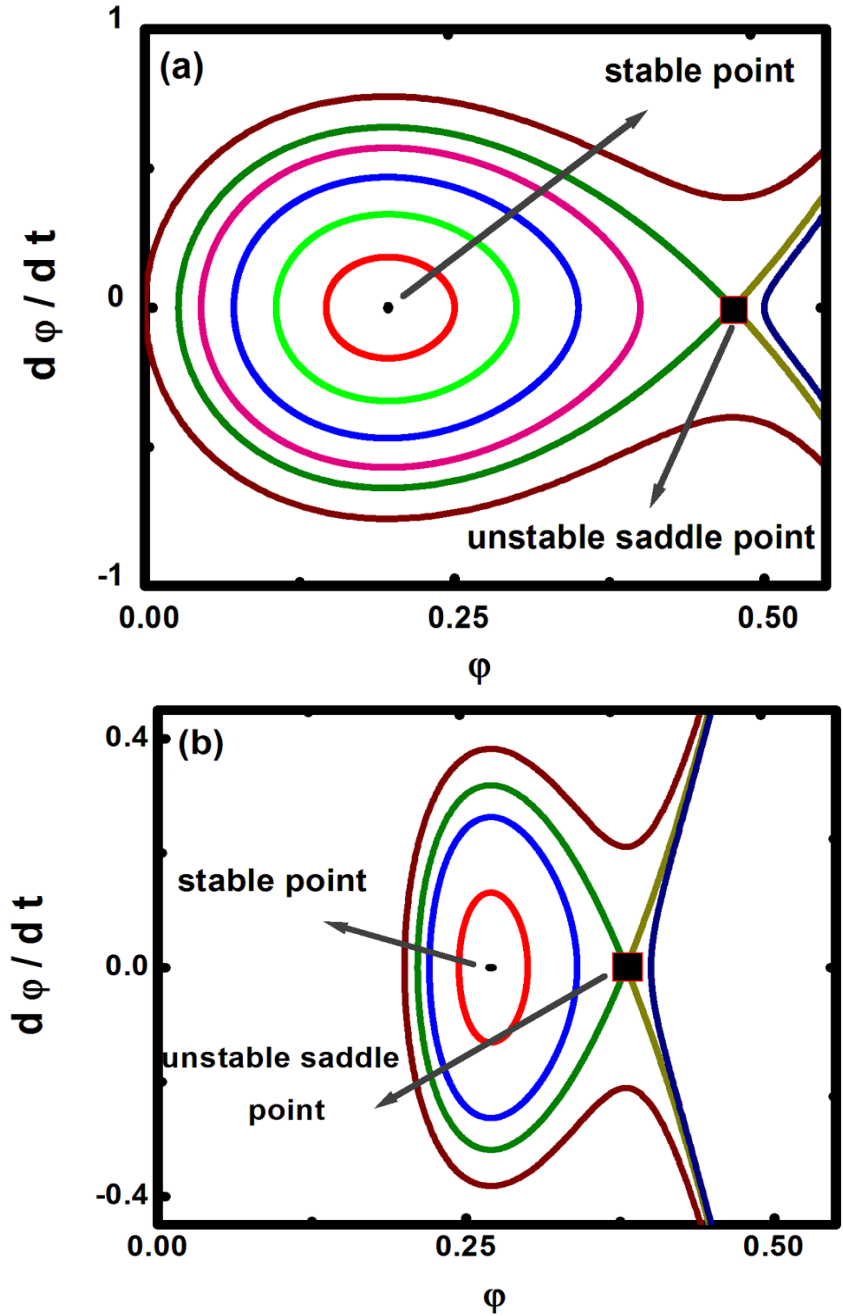

**Figure 7s** Phase portraits d$\varphi$/dt vs. $\varphi$ for $\delta_v = 0.05$ and $\delta_{Cas} = 0.1$, and initial conditions inside and outside of the homoclinic orbit. (a) Amorphous PCM, and (b) Crystalline PCM.

## 4. The effect of energy dissipation on the actuation of a torsional device

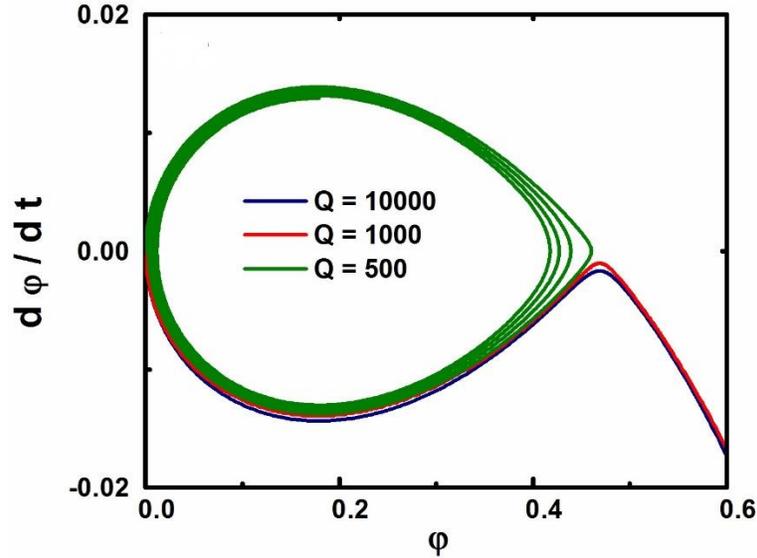

**Figure 8s** Influence of the damping term on actuation dynamics of torsional MEMS for the crystalline phase with $\delta_{Cas} = 0.1$, $\delta_v = 0$, and different values of the quality factor Q. A decreasing quality factor Q can change stiction to dissipative stable motion for torsional device.

**References**
[s1] E. M. Lifshitz. Sov. Phys. JETP **2,** 73 (1956); I. E. Dzyaloshinskii, E. M. Lifshitz and L. P. Pitaevskii. Sov. Phys. Usp. **4,** 153 (1961).
[s2] V. B. Svetovoy, P. J. van Zwol, G. Palasantzas, and J. Th. M. DeHosson, Phys. Rev. B. **77**, 035439 (2008).
[s3] G. Torricelli, P. J. van Zwol, O. Shpak, G. Palasantzas, V. B. Svetovoy, C. Binns, B. J. Kooi, P. Jost, and M. Wuttig, Adv. Funct. Mater. **22**, 3729 (2012).
[s4] G. Torricelli, P. J. van Zwol, O. Shpak, C. Binns, G. Palasantzas, B. J. Kooi, V. B. Svetovoy, and M. Wuttig, Phys. Rev. A **82**, 010101 (R) (2010).
[s5] M. Sedighi, V. B. Svetovoy, W. H. Broer, and G. Palasantzas, Phys. Rev. B **89**, 195440 (2014); M. Sedighi, V. B. Svetovoy, and G. Palasantzas, Phys. Rev. B **93**, 085434 (2016).